\documentclass[journal=jacsat,manuscript=article]{achemso}

\usepackage[version=3]{mhchem} 
\usepackage{xcolor}
\usepackage{siunitx}

\DeclareSIUnit\bar{bar}
\DeclareSIUnit\angstrom{\text {Å}}
\usepackage[utf8]{inputenc}
\DeclareUnicodeCharacter{00B7}{\cdot}

\author{W. Li}
\affiliation[LMU]
{Department of Physics, Ludwig-Maximilians-Universit\"{a}t Munich, D-85748 Garching, Germany}
\alsoaffiliation[MPQ]
{Max Planck Institute of Quantum Optics, D-85748 Garching, Germany}
\author{A. Saleh}
\affiliation[KSU]
{Attosecond Science Laboratory, Physics and Astronomy Department, King Saud University, Riyadh 11451, Saudi Arabia}
\author{M. Sharma}
\author{M. Sierka}
\affiliation[Jena]
{Otto Schott Institute of Materials Research, Friedrich Schiller University Jena, D-07743 Jena, Germany.}
\author{C. H\"{u}necke}
\affiliation[Jena2]
{Institute of Physical Chemistry, Friedrich Schiller University Jena, D-07743 Jena, Germany.}
\author{M. Neuhaus}
\author{L. Hedewig}
\author{B. Bergues}
\affiliation[LMU]
{Department of Physics, Ludwig-Maximilians-Universit\"{a}t Munich, D-85748 Garching, Germany}
\author{M. Alharbi}
\author{A. M. Azeer}
\affiliation[KSU]
{Attosecond Science Laboratory, Physics and Astronomy Department, King Saud University, Riyadh 11451, Saudi Arabia}
\author{S. Gr\"{a}fe}
\affiliation[Jena2]
{Institute of Physical Chemistry, Friedrich Schiller University Jena, D-07743 Jena, Germany.}
\alsoaffiliation[Jena3]
{Institute of Physical Chemistry, Institute of Applied Physics, and Abbe Center of Photonics, Friedrich Schiller University Jena, D-07743 Jena, Germany.}
\alsoaffiliation[Jena\_IOF]{Fraunhofer Institute of Applied Optics and Precision Engineering, D-07745 Jena, Germany}
\author{M. F. Kling}
\affiliation[LMU]
{Department of Physics, Ludwig-Maximilians-Universit\"{a}t Munich, D-85748 Garching, Germany}
\alsoaffiliation[MPQ]
{Max Planck Institute of Quantum Optics, D-85748 Garching, Germany}
\alsoaffiliation[SLAC]
{SLAC National Accelerator Laboratory, Menlo Park, CA 94025, USA}
\alsoaffiliation[Stanford]
{Applied Physics Department, Stanford University, Stanford, CA 94305, USA}
\author{A. F. Alharbi}
\affiliation[KSU]
{Attosecond Science Laboratory, Physics and Astronomy Department, King Saud University, Riyadh 11451, Saudi Arabia}
\alsoaffiliation[KACST]
{King Abdulaziz City for Science and Technology (KACST), Riyadh 11442, Saudi Arabia}
\author{Z. Wang}
\affiliation[LMU]
{Department of Physics, Ludwig-Maximilians-Universit\"{a}t Munich, D-85748 Garching, Germany}
\alsoaffiliation[MPQ]
{Max Planck Institute of Quantum Optics, D-85748 Garching, Germany}
\email{zilong.wang@physik.uni-muenchen.de}

\title[sHHG paper]
  {Resonance effects in Brunel harmonic generation in thin film organic semiconductors}

\abbreviations{OSCs,HHG}

\begin{document}







\begin{abstract}

Organic semiconductors have attracted extensive attention due to their excellent optical and electronic properties. Here, we present an experimental and theoretical study of Brunel harmonic generation in two types of porphyrin thin films: tetraphenylporphyrin (TPP) and its organometallic complex derivative Zinc tetraphenylporphyrin (ZnTPP). Our results show that the $\pi$-$\pi^\ast$ excitation of the porphyrin ringsystem plays a major role in the harmonic generation process. We uncovered the contribution of an interband process to Brunel harmonic generation. In particular, the resonant ($S_0 \rightarrow S_2$ transition) enhanced multiphoton excitation is found to lead to an early onset of non-perturbative behavior for the 5th harmonic. Similar resonance effects are expected in Brunel harmonic generation with other organic materials.

\end{abstract}

\section{Introduction}
Organic semiconductors (OSCs) are a group of semiconducting molecular materials, including small molecules and conjugated polymers, with quite different electronic structures from inorganic solids\cite{ahmad_organic_2014,bronstein_role_2020,okamoto_robust_2020}. The mechanism of charge transport and carrier dynamics in OSCs is very special, including two dynamics referred to as the intra- or inter-molecular processes \cite{coropceanu_charge_2007,jaiswal2006polymer}. Intra-molecular charge transport is based on delocalized $\pi$-electrons over the conjugated system, while inter-molecular charge transport depends on many factors such as structural arrangement, degree of order, density and temperature.
Thanks to the large freedom in the synthesis of materials, the possibility of tailoring optical and electronic properties and the ease of solid-state thin film fabrications, they have been found with many successful applications in flexible displays with organic light emitting diodes (OLED), photovoltaics, and field-effect transistors \cite{capelli2010organic,meng2018organic,yan2009high}. Additionally, they exhibit significant nonlinear properties, and are proposed for applications in saturable absorption, optical switching, and frequency up-conversion \cite{semin2021nonlinear}. 

High order nonlinear optical processes have been subject of extensive investigations in a variety of material systems. For instance, high harmonic generation (HHG) in gas-phase atoms and molecules is widely studied and may lead to the generation of attosecond extreme ultraviolet pulses used in attosecond time-resolved spectroscopy \cite{krausz_attosecond_2009,li_attosecond_2020}. More recently, such studies were extended to inorganic solids, including bulk \cite{ghimire_observation_2011,you_anisotropic_2017} and low-dimensional crystals \cite{yoshikawa_high-harmonic_2017,tancogne-dejean_atomic-like_2018}. The related work provided insight into light field-driven electron dynamics in these systems, and their potential application as solid-state UV sources \cite{ghimire_high-harmonic_2019,you_high-harmonic_2017}. While OSCs bear potential for high-frequency up-conversion in integrated optical and photonic devices, the high-order processes in solid-state OSCs remain to be studied.

Porphyrins and their derivatives, i.e. metalloporphyrins, are one of the most studied OSC systems due to their excellent optical \cite{senge_nonlinear_2007} and optoelectronic properties\cite{umeyama2009synthesis}. In particular, the insertion of different types of metal atoms into the center of the porphyrin ring modifies the electronic structures and nonlinear properties in metalloporphyrins compared to the metal-free ones \cite{nurhayati_temperature-dependent_2020, kielmann_molecular_2019}. For example, the Zn-based metalloporphyrins exhibit higher two-photon absorption coefficients than the metal-free ones \cite{cassagne2015nonlinear}. In this work, we experimentally investigated and compared Brunel harmonic generation below the ionization potential ($I_p$) in tetraphenylporphyrin (TPP) and Zinc tetraphenylporphyrin (ZnTPP) thin films excited by strong infrared ultra-short laser pulses. The experimental results are interpreted with theoretical calculations based on real-time - time-dependent density functional theory (RT-TDDFT). Herein, we utilize two slightly different variants or implementations: the newly-developed variant based on localized (atomic) orbitals as very recently implemented in the Turbomole program package\cite{RT-TDDFT_Turbomole_2020} and the commonly successfully employed variant employing real-space grids as in the Octopus program package\cite{Castro2015, Octopus_2015}. Our results show that resonant and non-resonant transitions within the ZnTPP and TPP molecules occur. In particular, an efficient transition to the $\pi$-$\pi^\ast$ excited states of the porphyrin ringsystem (accessible by a 5-photon transition) changes the nonlinear field-induced dynamics substantially. These resonances change the character of the corresponding harmonic emission substantially, manifesting itself, e.g., by a different slope in the intensity scaling.

\section{Methods and sample preparation}

\begin{figure}[ht]
  \includegraphics[width=16cm]{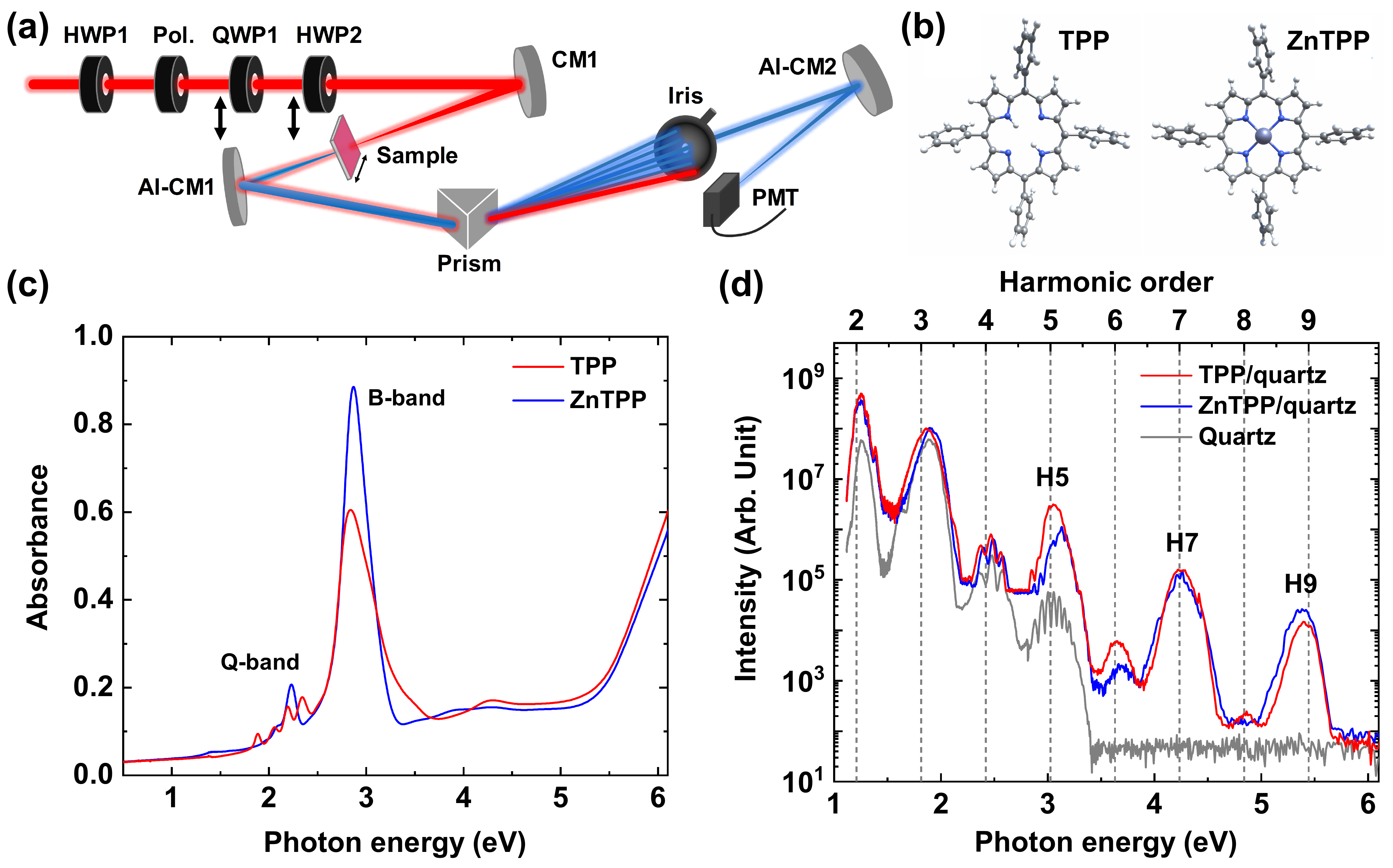}
  \caption{(a) Experimental setup; (b) Molecular structure of TPP (left) and ZnTPP (right). (c) Linear absorption spectra of 100-nm thick TPP and ZnTPP films; (d) High-order harmonic spectra of 100-nm thick TPP (red, solid line) and ZnTPP (blue, solid line) films as well as the bare quartz substrate (grey, solid line)}
  \label{fgr:Figure 1}
\end{figure}

The experimental setup for the below-threshold harmonic generation is shown in Fig. \ref{fgr:Figure 1}(a). Samples were excited with \SI{2}{\micro\meter} laser pulses from an optical parametric chirped pulse amplification (OPCPA) system, with a typical pulse duration of $\sim$ \SI{30}{\femto\second} (full width at half maximum, FWHM) and a repetition rate of \SI{100}{\kilo\hertz} (see supporting information, SI)\cite{neuhaus_10_2018}. A half ($\lambda /2$) waveplate (HWP1) and a wire grid polarizer were used to vary the laser intensity. An additional  quarter ($\lambda /4$) waveplate and a second half waveplate (HWP2) were added in the beam path after the HWP1 and polarizer in order to control and vary the laser ellipticity while keeping the direction of the major axis of the elliptical polarization fixed at all times. 
Transmitted harmonic emission was collected and focused onto a photomultiplier tube (PMT, Hamamatsu) by two UV-enhanced aluminum concave mirrors (Al-CM1 and Al-CM2). The signal of each order of harmonics was selected by a combination of CaF$_{2}$ equilateral prism and an iris placed in front of the PMT. The measurement of harmonic emission spectra as well as the calibration of prism rotation angle were done with a UV-VIS spectrometer (Ocean Optics HDX). To avoid the sample degradation under high excitation intensity, a mechanical shutter was placed in front of the sample and was switched on/off between each laser excitation; the signal integration time was kept as low as possible (\SI{100}{\milli\second}). Furthermore, the sample was mounted on a two-axis stage, shifting the sample position after each single measurement. The entire measurement setup was enclosed in a home-built purge box filled with nitrogen gas. In our experiment, the same measurements were repeated more than three times on different batches of samples, showing the reproducibility of the experimental results.

TPP and ZnTPP (from Aldrich Chem Co., purity $\geq 99\%$), as the investigated OSC materials, were used as received without any further purification. Molecular structures of TPP and ZnTPP can be seen in Fig. \ref{fgr:Figure 1}(b), where the (deposited) TPP molecule has $D_{2h}$ symmetry and the ZnTPP molecule has $D_{4h}$ symmetry with the Zn fitted in the center of the planar tetrapyrrolic ring system. \cite{zeyada2016structural,marsh1996microscale} Thin-film samples were prepared with the physical vapor deposition technique on optically flat quartz substrates in a vacuum coating system (Edwards, E306 A) under a base pressure of $10^{-6}$ \si{\milli\bar}, yielding a typical film thickness of \SI{100}{\nano\meter}. All the organic thin films for measurements were as-deposited samples without any further treatments. Surface topography images of both samples measured by atomic force microscopy (AFM) (see Fig. S1 in SI) shows a homogeneous distribution of sample materials under regions of study. Both of the thin-film samples consist of randomly oriented nanorod-shape molecular packing structures dispersed in the amorphous matrix, where both parallel molecular stacking and monomer can be found in the film \cite{zhang2016identifying}.

Linear absorption spectra of the produced TPP and ZnTPP thin films are shown in Fig. \ref{fgr:Figure 1}(c). Both samples exhibit the prominent B (Soret) band (\SI{2.84}{\eV} for TPP, \SI{2.87}{\eV} for ZnTPP) and a weak Q-band absorption corresponding to $S_0 \rightarrow S_2$ and $S_0 \rightarrow S_1$ $\pi-\pi^\ast$ singlet transitions, respectively. The TPP thin film shows four characteristic peaks (\SI{1.88}{\eV}, \SI{2.06}{\eV}, \SI{2.19}{\eV}, \SI{2.34}{\eV}) in Q-band region, while they collapse into two peaks (\SI{2.09}{\eV}, \SI{2.23}{\eV}) due to the higher symmetry ($D_{4h}$) of ZnTPP. \cite{zeyada2016structural,taniguchi2021comprehensive,marsh1996microscale}

Theoretical calculations were done with two different implementations of RT-TDDFT. The first set of calculations is based on our recently developed implementation of the RT-TDDFT based on localized atomic basis functions \cite{RT-TDDFT_Turbomole_2020} within the Turbomole program suite\cite{Turbomole_2020} (the Turbomole implementation hereafter). The localized basis method is able to handle large molecular systems, such as the TPP molecule, with high computing efficiency ($\sim 10^3$  CPU hours/calc); however, employing atomic-centered basis functions, ionization and field-driven continuum electronic dynamics outside the molecule cannot be addressed. For the here investigated harmonic generation with energies of quanta below the ionization threshold, it is generally well suited. A low or even negligible level of ionization is confirmed experimentally, but also with the second variant of simulation based on the well-established real-space grid-based Octopus program package (the Octopus implementation hereafter).\cite{Octopus_2015}. Many works have demonstrated the success of this method to simulate high-harmonic generation in molecules or bulk.\cite{tancogne-dejean_atomic-like_2018} Different from the Turbomole implementation, electronic wavefunction and current are calculated in real space in a spherical simulation box with a radius of \SI{18}{\angstrom}, therefore, field-driven electronic dynamics (and, thus, also possible field-driven re-combination, giving possibly rise to higher-order harmonic generation) following ionization are included.

\section{Results and discussion}

The measured harmonic emission spectra of TPP (red) and ZnTPP (blue) thin film samples excited at an intensity of \SI{2.18}{\tera\W/\centi\meter^2} are shown in Fig. \ref{fgr:Figure 1}(d). The 5th (H5, \SI{3.1}{\eV}), 7th (H7, \SI{4.25}{\eV}) and 9th (H9, \SI{5.4}{\eV}) order harmonic emissions from both organic thin film samples can be clearly distinguished in the spectra. Emission signatures from the quartz substrate (grey) were recorded as a reference, whose contributions are seen only up to 4th order. For both OSC samples, the H5 peaks are practically on-resonance with the B-band absorption. Weak even-harmonic emission (6th and 8th) can be seen from samples arising from the partial breaking of inversion symmetry in the polycrystalline thin film samples with random orientation of individual molecules. The last observed harmonic, i.e. H9, which lies at \SI{5.4}{\eV} below the ionization thresholds of the two molecules, which are reported as \SI{6.32}{\eV} for TPP and \SI{6.06}{\eV} for ZnTPP. \cite{experimental_ionization_potentials_1976} This, together with the fact, that we do not observe damage of the samples, see above, indicates that the Corkum-type harmonic mechanism for atoms and molecules, involving ionization, acceleration and recombination does not play the dominant role, as we exclusively see harmonic below the ionization threshold.

\begin{figure}[t!]
  \includegraphics[width=16cm]{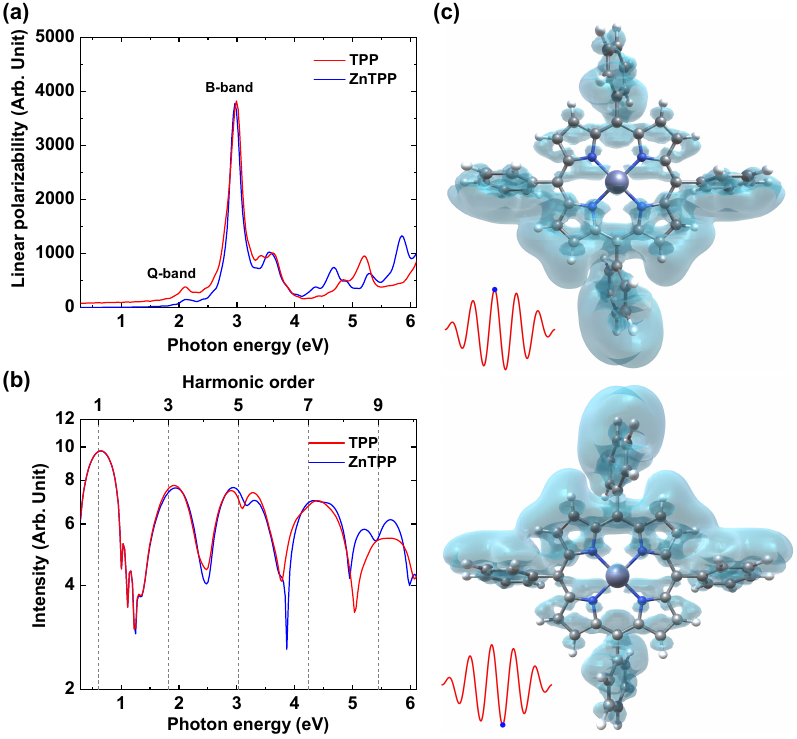}
  \caption{(a) Linear absorption spectra and (b) harmonic spectra of TPP and ZnTPP single molecules obtained by Turbomole implementation; (c) Isosurface snapshots of the difference between excited and ground state electronic density in ZnTPP single molecule at the maximum and minimum of the linearly polarized laser pulse (visualization via CrysX-3D Viewer\cite{CrysX_2019} with isovalue $=+7\times 10^{-5}e/\si{\angstrom^3}$; also see the SI for the complete animation)}
  \label{fgr:Figure 2}
\end{figure}

The harmonic emission spectra of both samples were calculated and compared using the above mentioned RT-TDDFT simulations. For most calculations, there is almost no difference between the two computational tools (see SI for complete numerical results). We therefore concentrate only on the results from the computationally more efficient implementation, i.e. the Turbomole. For those cases where the numerical results differ, we will evoke both results and explain the origin of the differences. We first calculate the linear absorption spectra of the single-molecule TPP and ZnTPP. As can be seen in Fig. \ref{fgr:Figure 2}(a), the experimentally measured spectra of thin film samples and the calculated spectra of the individual molecules agree very well, particularly in terms of the B-band absorption. Thus, in what follows, we will remain in the molecular picture (rather than evoking the bulk perspective) for the description of the harmonics generation.

Harmonic generation processes were simulated by interaction of a \SI{2}{\micro\meter} linearly polarized cosine/trapezoid-shaped laser pulse with 5 optical cycles ($\sim$ \SI{33}{\femto\second}) and a field intensity of \SI{1.5}{\tera\watt/\centi\meter^2}, similar to the experimental conditions, with the corresponding molecules. For the simulations, we first calculated the equilibrium structure of the individual molecule in vacuum. Subsequently, starting from the electronic ground-state density, the system interacts with the above-defined laser pulses, which drive the electron dynamics. To make sure the Turbomole method is within the scope of application under such conditions, the ionization potentials and overall level of ionization of the molecules are calculated using Turbomole and Octopus, respectively. The ionization potentials of the TPP and ZnTPP molecules, as calculated via Turbomole, are \SI{6.47}{\eV} and \SI{6.51}{\eV}, respectively, which are in reasonable agreement with the measured values of \SI{6.32}{\eV} for TPP and \SI{6.06}{\eV} for ZnTPP \cite{experimental_ionization_potentials_1976} and higher in energy than H9. Furthermore, The overall ionization estimated from the loss of charges due to absorbing boundary conditions is very low ($< 0.01\%$) for the experimental conditions. This is also true from experimental point of view, where sample degradation induced by the laser pulses was carefully avoided. The above two findings lend confidence in the ability of Turbomole to be suitable for this study and therefore, the theoretical analysis is continued with the method utilizing local basis sets, the Turbomole program package, as it is computationally much more efficient.

Harmonic emission spectra of TPP and ZnTPP molecules obtained from the Turbomole calculations are shown in Fig. \ref{fgr:Figure 2}(b). They are obtained by Fourier transforming the time-dependent dipole acceleration, $\ddot{\mu}(t)$, driven by the few-cycle laser field (see also in \ref{fgr:Figure 2}(c)). The spectra exhibit almost identical spectral shape for both molecules except for the H5 and H9. Such identical spectra, particularly in terms of the harmonic intensity from the two molecules, indicate similar electronic dynamics within the laser pulses. Indeed, the snap-shot images of electron wavepacket distributions at different times in Fig. \ref{fgr:Figure 2}(c) reveal that the excitation is dominated by contribution from the $\pi$ and $\pi^\ast$ orbitals. The contribution of the central Zn atom's d-orbitals to the $\pi^\ast$ orbitals of the surrounding TPP ring is minor or even negligible. Emission band splitting is observed in H5 at $\sim$ \SI{3}{\eV}, which can be attributed to the on-resonant $\pi\rightarrow\pi^{*}$ excitation at around \SI{2.99}{\eV} (see absorption spectra in Fig. \ref{fgr:Figure 2}(a)). This is not seen experimentally, probably due to propagation effects, focal volume averaging, the random molecular orientation in the thin film samples, as well as resonance-enhanced multiphoton excitation and possibly ionization processes, which are not included in the modeling.

\begin{figure}[ht]
  \includegraphics[width=16cm]{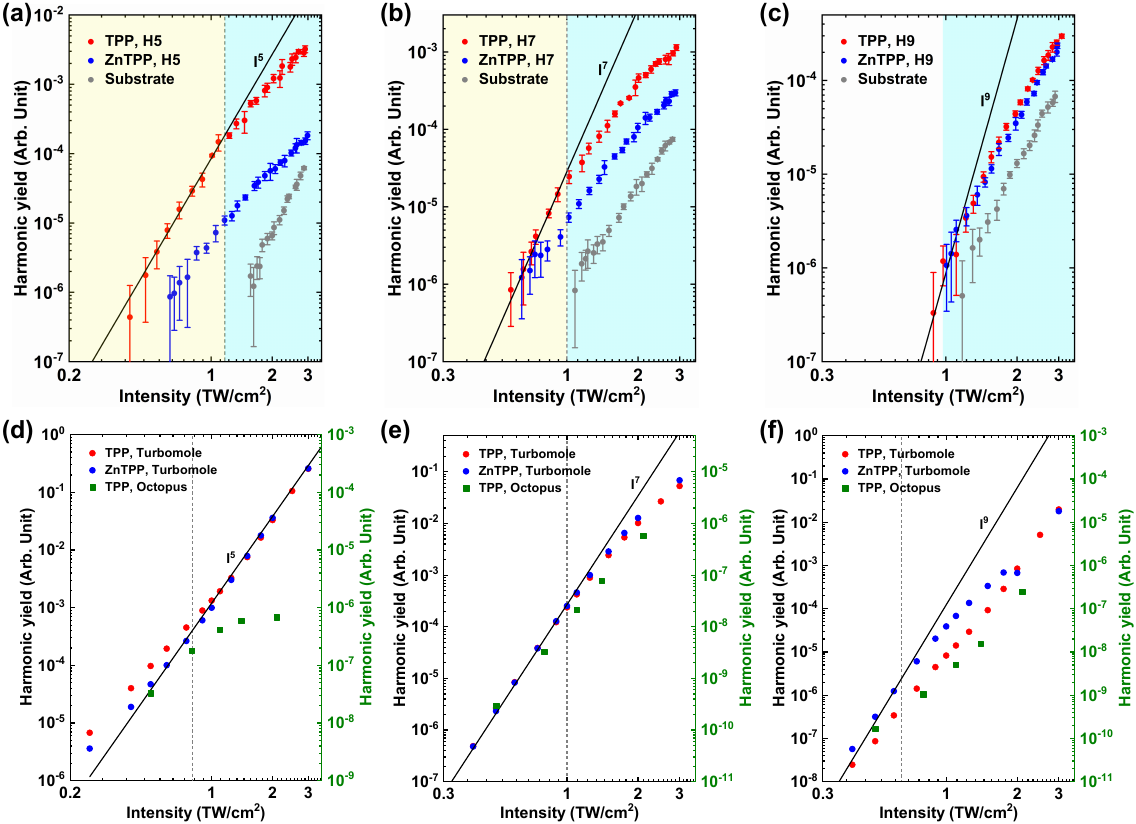}
  \caption{Excitation intensity dependence characteristics of harmonic signals in TPP and ZnTPP. Experimental results from 100 nm thick thin films: (a) H5; (b) H7; (c) H9 (error bar: standard deviation of 400 measurements under the same conditions); theoretical simulation results obtained with Turbomole (red and blue dots) and Octopus (olive squares): (d) H5; (e) H7; (f) H9}
  \label{fgr:Figure 3}
\end{figure}

In order to obtain mechanistic insight into the origin of harmonic emission, we studied the excitation intensity dependence for each harmonic, as shown in Fig. \ref{fgr:Figure 3}. Measured harmonic yields for H5, H7 and H9 from both TPP (red) and ZnTPP (blue) thin films are shown in Fig. \ref{fgr:Figure 3}(a-c). The grey scatters in the figures represent background noise including signals contributed from emissions in quartz substrate as well as stray light entering into the PMT detector. Harmonic emission from TPP sample is significantly higher than that from ZnTPP sample except for H9 emission. The perturbative light-matter interaction regime can be immediately identified, marked as yellow-shaded area, from the $Y \propto I_{exc}^{n}$ dependency in H5 and H7 (Fig. \ref{fgr:Figure 3}(a-b)), where $I_{exc}$ is excitation laser intensity and $n$ is the harmonic order ($n$ = 5, 7). For H5 and H7, perturbative behavior dominates up to intensities of about \SI{1.2}{\tera\W/\centi\meter^2} and \SI{1}{\tera\W/\centi\meter^2}, respectively. The deviation of $I_{exc}^{n}$ at higher intensity (blue-shaded region) indicates the transition from perturbative to non-perturbative interaction in the samples. For H9, only non-perturbative light-matter interaction can be seen in Fig. \ref{fgr:Figure 3}(c). We do not reach intensities where the signal is saturated, as this would lead already to sample damage. The Keldysh parameter, $\gamma=\sqrt{I_p/2U_p}$ at \SI{3}{\tera\W/\centi\meter^2} in the non-perturbative regime, where $U_p$ is the pondermotive energy, is calculated as 1.60 and 1.56 for TPP and ZnTPP, respectively, which indicates that both multiphoton and tunneling ionization occur in this regime \cite{wolter2015strong,kruchinin2018colloquium}. The exceptionally larger yield for H5 from TPP compared to that from ZnTPP is attributed to the re-absorption of the emitted harmonic upon propagation through the thin film samples. This onset of deviation of the perturbative character already by comparatively low intensities for H5 may be at first surprising and will be explained in what follows. The simulated excitation intensity-dependent emission was obtained by integrating the spectral peaks of each order at each excitation intensity, shown in Fig. \ref{fgr:Figure 3}(d-f). The $Y \propto I_{exc}^{n}$ dependency at relatively low excitation could be well reproduced from both theoretical implementations, further confirming that the observed harmonic emissions are due to the perturbative nonlinear behavior of the molecule. Deviation from perturbative interaction regime can be seen in H7 and H9 from both simulation methods. Interestingly, large differences between the model simulations are seen for H5: Turbomole predicts perturbative behavior for H5 throughout the intensity range while Octopus results show a saturation behavior after around \SI{0.8}{\tera\W/cm^2}, in better agreement with the experimental observation. This is because the Octopus implementation calculates electron wavepackets on real-space grid basis, where electron ionization and the subsequent light-driven dynamics as well as recombination processes are included. 
In fact, a population analysis (within Octopus) shows that upon interaction with the driving field, the electronic ground state is strongly depleted. The B-band is thereby populated efficiently. However, with the B-band populated, transitions into multiply other electronic states, either bound or even transiently virtual states occur, thereby changing the nonlinearity substantially. This observation marks the transition from the molecular-type behavior towards the bulk where electronic dynamics in bands occur, formed by delocalized and degenerate electronic states of the same electronic character. A similar scenario occurs here for the large molecules, despite the obvious difference that we do not have periodic structures. Therefore, despite of the calculated rather low ionization rate in the molecules, the inclusion of ionization processes that electrons leave the molecule far away and subsequently recombined driven by the laser field plays an important role in the non-perturbative behavior of H5 emission at high excitation intensities.

\begin{figure}[ht]
  \includegraphics[width=11cm]{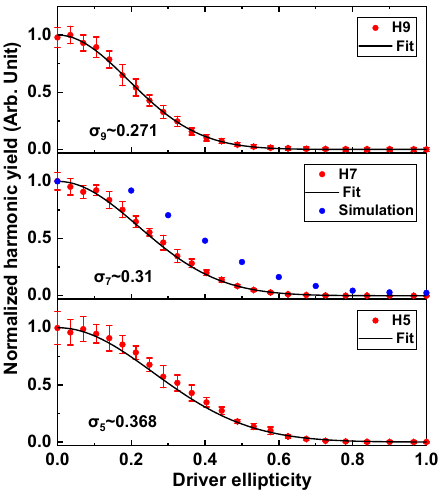}
   \caption{Excitation ellipticity dependence characteristics of high harmonic signals in 100-nm thick TPP: normalized high harmonic yields (red dots) versus driver ellipticity, a Gaussian function fitting (black solid line) and simulation results obtained from Turbomole implementation (blue dots)}
  \label{fgr:Figure 4}
\end{figure}

An additional measurement is the excitation ellipticity dependence of the Brunel harmonic emission which is investigated and shown in Fig. \ref{fgr:Figure 4}. Normalized yields of H5, H7 and H9 under different driver ellipticities in the non-perturbative regime from TPP thin film samples are demonstrated in the figure. Results from ZnTPP samples exhibiting the same trend as TPP, therefore, are only shown in SI. In the experiment, the driver ellipticity, \(\varepsilon_d\), of exciting laser pulses is varied from 0 (linearly polarized) to 1 (circularly polarized) while keeping the total intensity and major axis fixed all the time. As expected, for all harmonics, the maximum yield occurs for linearly polarized excitation while nearly no signal can be measured for circularly polarized excitation. Higher order harmonics, e.g. H9, show narrower dependence curve shape than lower order ones, indicating a higher sensitivity to the change in excitation ellipticity. The ellipticity-dependent harmonic emission behavior can be well fitted by the Gaussian function \(e^{-\varepsilon_d^2/\sigma_n^2}\), where \(\varepsilon_d\) is the driver ellipticity and \(\sigma_n\) (n=5,7,9) is fitting parameter\cite{liu_effect_2016}. A lower \(\sigma_n\) represents the higher sensitivity of harmonic yield to the excitation ellipticity. This behavior can be explained by the increase of non-perturbative character from H5 to H9. Moreover, the much narrower width of H9, which lies energetically only a little below the ionization threshold, points towards a higher contribution of (transient) ionization to the signal. The relatively broad ellipticity dependence of H5 can be explained by the role of the resonance as well: as the transition of the $\pi\rightarrow\pi^{*}$ state is dipole-allowed within the plane, for almost all ellipiticities, the electronic state can be coherently excited, thereby driving a wavepacket. Similar behavior has also be reported for the HHG in plateau regime (above-$I_p$) in gaseous atoms and small molecules where the recombination probability of the ionized electron with its parent ion is greatly reduced in elliptically polarized electric fields\cite{shan_effect_2004,kanai_ellipticity_2007,Alharbi_Effects_2017}. Additionally, the measured ellipticity dependence still features a narrower peak than that of the simulated results by Turbomole on H7, shown in Fig. \ref{fgr:Figure 4}, suggesting further implying that the atomic-like transient ionization processes should also involve in the harmonic generation processes for other harmonic orders. Not surprisingly, we note that measured ellipticity dependence behavior does not fit with the total harmonic yield generated by the two electric field components orthogonally decomposed from elliptical polarized excitation laser field.

\section{Conclusion}
We have demonstrated Brunel harmonic generation in TPP and ZnTPP thin films both in perturbative and strong-field driven non-perturbative regime. With the help of RT-TDDFT based calculations using Turbomole and Octopus methods, we identify the origin of unusual scaling of the 5th harmonic to be caused by excitaiton to a bright, resonant electronic state (the B-band) of the molecule, which changes the underlying dynamics substantially: first of all, the 5-photon resonant state causes a strong depletion of the electronic ground state of the molecule with markable population of this excited state and large contribution of many other bound (and quasi-free transient) excited electronic states. This results in a strong deviation from the perturbative mechanism of harmonic generation, seen in the intensity scaling but also the ellipticity dependence. The B-band is energetically so broad that excitation into it leads to a dynamics resembling bulk-like behaviour. In addition, upon light propagation through the sample, it can be efficiently re-absorbed and re-emitted, changing the observed features even further. In contrast, the 9th harmonic lies energetically closely around the ionization threshold. Thus, in order to fully describe its mechanistic details, partial (transient) ionization cannot be neglected any more. Thus, the Brunel harmonic generation mechanism originating from excitation nonlinearities seems to be the dominant mechanism of harmonic generation in these systems. Our studies show the potential of organic semiconductors, TPP and ZnTPP in our case, as high frequency up-conversion emitters. And they are applicable integrated optical photonic and opto-electronic devices thanks to the mature device fabrication techniques.

\begin{acknowledgement}


We are grateful for support by King-Saud University within the MPQ-KSU collaboration. W.L. acknowledges support by the Max Planck Society via the IMPRS for Advanced Photon Science. M.F.K. acknowledges support by the Max Planck Society via the Max Planck Fellow program. M.F.K.'s work at SLAC is supported by the U.S. Department of Energy, Office of Science, Basic Energy Sciences, Scientific User Facilities Division, under Contract No. DE-AC02-76SF00515. Z.W. is grateful for support by a startup grant from the SPP QUTIF funded by the German Research Foundation, and for support by the Alexander von Humboldt Foundation. M.S. and M.S. gratefully acknowledge financial support from the Turbomole GmbH as well as the German Research foundation DFG (CRC 1375 NOA, project number 398816777) project A4. S.G., C.H., M.S. and M.S. are grateful for support by the German Research Foundation DFG (CRC 1375 NOA, project number 398816777) projects A1, A4 and B1.

\end{acknowledgement}

\begin{suppinfo}

TPP and ZnTPP thim films preparation and characterization; detailed experimental and simulation procedures; additional experimental and theoretical results.

\end{suppinfo}

\bibliography{Manuscript}

\end{document}